\documentclass[
    ,final            
  ]
  {aipproc}
\usepackage{amssymb,latexsym,amscd,amsmath,epsfig}
\usepackage{longtable}

\def\be{\begin{equation}}
\def\ee{\end{equation}}
\def\ba{\begin{array}}
\def\ea{\end{array}}
\def\bea{\begin{eqnarray}}
\def\eea{\end{eqnarray}}
\def\bi{\begin{itemize}}
\def\ei{\end{itemize}}

\layoutstyle{8x11single}

\begin{document}

\title{Modeling Nuclear Pasta and the Transition to Uniform Nuclear Matter with the 3D-Skyrme-Hartree-Fock Method}


\author{W.~G.~Newton}{
  address={Department of Physics, Texas A\&M University-Commerce, Commerce, Texas 75429-3011, USA}
}

\begin{abstract}
The first results of a new three-dimensional, finite temperature Skyrme-Hartree-Fock+BCS study of the properties of inhomogeneous nuclear matter at densities and temperatures leading to the transition to uniform nuclear matter are presented. A constraint is placed on the two independent components of the quadrupole moment in order to self-consistently explore the shape phase space of nuclear configurations. The scheme employed naturally allows effects such as (i) neutron drip, which results in an external neutron gas, (ii) the variety of exotic nuclear shapes expected for extremely neutron heavy nuclei, and (iii) the subsequent dissolution of these nuclei into nuclear matter. In this way, the equation of state can be calculated across phase transitions from lower densities (where one dimensional Hartree-Fock suffices) through to uniform nuclear matter without recourse to interpolation techniques between density regimes described by different physical models.
\end{abstract}

\maketitle


\section{Introduction}

The equation of state (EoS) of nuclear matter is one of the main theoretical contributions to models of neutron stars (NSs) and collapsing stellar cores (SNe). It is derived from the microphysics of nucleon-nucleon interactions in-medium and is ill-constrained throughout most of the density and temperature regime realised in astrophysical situations (baryon number densities $n_{\rm b} \sim 10^{-11}$ fm$^{-3}$ to $\sim$1 fm$^{-3}$ and temperatures $\sim 0 - 100$MeV). As such, we look to observational properties of NSs and SNe to serve as constraints on the bulk properties of nuclear matter and its EoS.

To facilitate this endeavor, the EoS of stellar matter should be constructed, as far as possible, using the same underlying physical model and at the same level of approximation over the whole range of densities and temperatures occuring in SNe and NSs. The EoS should be calculated self-consistently across all relevant phase transitions and where multiple phases co-exist. Quantities that are specified with a given EoS (e.g. pressure, energy density) should
be extended to include electrical, mechanical, thermal and transport properties of matter (as has been argued in Lin et al~\cite{Lin2007}), again, all calculated with the same underlying physical model and at the same level of approximation.

An important phase transition in bulk nuclear matter occurs at densities just below nuclear saturation density $n_{\rm s} \approx 2.5 \times 10^{14}$ g cm$^{\rm -3}$ when inhomogeneous nuclear matter (nuclei or nuclear clusters immersed in a neutron gas) dissolve into uniform nuclear matter. At densities just below the transition density the inhomogeneous matter is frustrated and may undergo a series of transitions in the shape of the nuclear clusters; from spherical to cylindrical to slab to cylindrical hole to spherical hole, commonly referred to as the nuclear `pasta' phases~\cite{rave83,hashi84,hashi84_2}. Developing an EoS across this transition region together with a self-consistent determination of the transition density at various temperatures is a challenging task; in this proceeding a new attempt to do just this will be discussed.

There have been two methods employed to describe inhomogeneous nuclear matter self-consistently from a microscopic stand-point (using nucleons as the fundamental degrees of freedom). The Quantum Molecular Dynamics (QMD) method and similar~\cite{maruy1998,horowitz2004_1,horowitz2004_2,watanabe01_2, watanabe_2_05} are semi-classical approaches which can treat large numbers of nucleons at once (up to $\sim$ 100,000 so far), thus describing long-wavelength thermal and transport effects; however they do not quantum mechanical shell effects and band structure self-consistently, if at all. The mean-field Hartree-Fock (HF) method~\cite{Bender2003} is fully quantum mechanical; however, computational limitations currently confine HF calculations to dealing with a few thousand nucleons at most. The two approaches must therefore be viewed as complementary to each other.

In this proceeding calculations performed using the HF method with the widely used Skyrme energy-density functional are discussed. In such calculations, one assumes that matter can be described as locally consisting of periodic unit cells containing one nucleus or nuclear cluster. At densities much lower than nuclear saturation density it is adequate to perform calculations in one dimension (spherical symmetry) using the spherical Wigner-Seitz (WS) approximation, in which the unit cell is replaced by a spherical cell of the same volume; however, as one approaches the transition density the nuclear clusters become more and more closely spaced and deformed, and the spherical WS approximation breaks down~\cite{Chamel2007}. Three-dimensional calculations are therefore performed in cartesian co-ordinate space. The 3D calculations are intended to act as a consistent bridge between the lower density 1D calculations and the uniform matter calculations for which the HF equations yield an analytic solution.

Our study extends work done by Magierski and Heenen and more recently by G\"{o}gelein and Muther~\cite{magie02, gogel07}. Both 3D-HF studies to date have calculated nuclear configurations at only a limited number of values of density, proton fraction, and number of nucleons in the unit cell, and only for neutron star crustal matter. Only the most recent~\cite{gogel07} has attempted a calculation at finite temperature. The study of Magierski and Heenen~\cite{magie02} imposed a constraint on one of the proton quadrupole moments; however, to self-consistently probe the energy of various pasta shapes, both independent quadrupole moments should be constrained. Finally, it is not clear what role the numerical method itself plays in the results obtained; for example, what are the effects of the finite computational cell?

The computational scheme presented in this paper is constructed to be flexible enough to easily adopt a variety of nuclear interactions other than those of the Skyrme type, and to study different boundary conditions. Notably, in the application to zero-temperature beta-equilibrium neutron star matter requires full Bloch boundary conditions to determine the band structure of the neutron gas single particle energy spectrum. In this proceeding some numerical and physical properties of the 3D-HF method applied to inhomogeneous matter are discussed in the light of the first results of our calculations.

\section{Numerical Details}

The Skyrme-Hartree-Fock (SHF) method is widely used and has been documented extensively elsewhere~\cite{Bender2003}. A full account of the details of the code used in this study can be found in~\cite{thesis}. Here some of the computational details relevant to the study of nuclear matter are highlighted.

The Hartree-Fock equations are solved in co-ordinate space in one cubic unit cell of the locally periodic nuclear matter. Note that this choice admits several different crystal lattice types; simple cubic (sc), body centered cubic (bcc) and face centered cubic (fcc). To reduce the computational task further, only nuclear configurations that conserve reflection symmetry in the three Cartesian directions are considered. It follows that the computation need be performed only in one octant of the unit cell. The spin-orbit interaction is currently omitted. Pairing is included at zero-temperature using the BCS formalism with a simple contact potential and finite temperature effects are imposed through a Fermi-Dirac distribution of single-particle occupation numbers. Periodic boundary conditions are enforce implicitly by using the Fourier transform method to calculate derivatives and the Coulomb potential.

Due to the frustrated nature of the matter, it is expected that the absolute minimum energy of the cell is not going to be particularly pronounced and there will be a host of local minima separated by relatively small energy differences. In order to systematically survey the `shape space' of all nuclear configurations of interest, the quadrupole moment of the neutron density distribution is parameterized, and those parameters constrained. It is expected that the proton density distribution will follow closely that of the neutrons, so no constraint is imposed on the proton density quadrupole moments. Deformations are referred to in terms of the deformation `polar' co-ordinates ($\beta, \gamma$). $\beta$ gives the magnitude of the deformation and $\gamma$ its direction, ranging from prolate shapes ($\gamma = 0^{\rm o}$) to oblate ($\gamma = 60^{\rm o}$) shapes.

In our description of both SN and NS matter, the only non-nucleonic component of matter present in our calculations is a uniform density background of electrons.

In the work presented, unless stated otherwise, the Skyrme parameterization SkM$^*$ \cite{Bartel1982} is used.

\section{Results}

A selection of results are presented to illustrate a range of numerical and physical effects encountered.

\subsection{\label{sec2c2} Numerical Considerations}

\paragraph{Effect of Finite Cell Size on Nuclear Shape}

The effect of the finite computational volume on results obtained must be carefully examined. Firstly, does the geometry of the cell affect the nuclear shapes obtained and their energies?

\begin{figure}[!t]
\hspace{1pc} \centerline{\psfig{file=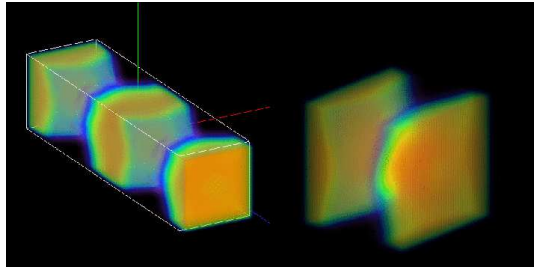,width=7cm}}
\caption{Obtaining a double nuclear shape. The right picture shows a 3D rendering of the neutron density profile at $n_{\rm b}$ = 0.08 fm$^{\rm -3}$, $y_{\rm p}$ = 0.3, $T$ = 0.0 MeV, $A$ = 700, and $(\beta,\gamma)$ = (1.0, 0$^{\rm o}$). The configuration shown in the left picture is the result of a calculation in a box double the size in the z-direction - with $A$ = 1400 - at otherwise the same parameter values. Blue indicates the lowest densities and red the highest.} \hspace{1pc} \label{Fig:1}
\end{figure}

Calculations were performed in a box that is double the length in the z-direction compared to the other two directions. The expectation was to find a configuration similar to the one in a cubic box at the same baryon number density (with half the number of nucleons), with the same free energy energy. Fig.~\ref{Fig:1} displays one such configuration. The nucleon density distribution has been rendered in three dimensions, with blue indicating the lowest densities and red the highest. The right-hand cell displays the nuclear shape obtained within a cubic cell; it is similar to the `lasagna' phase, with an additional cylindrical bridge joining the slabs. The left-hand cell displays a shape that is very similar to two instances of the modified lasagna shape adjacent to each other. The free energy densities of the two configurations agree to within 1 part in $10^{\rm 4}$.

\begin{figure}[!b]
\hspace{1pc}
\centerline{\psfig{file=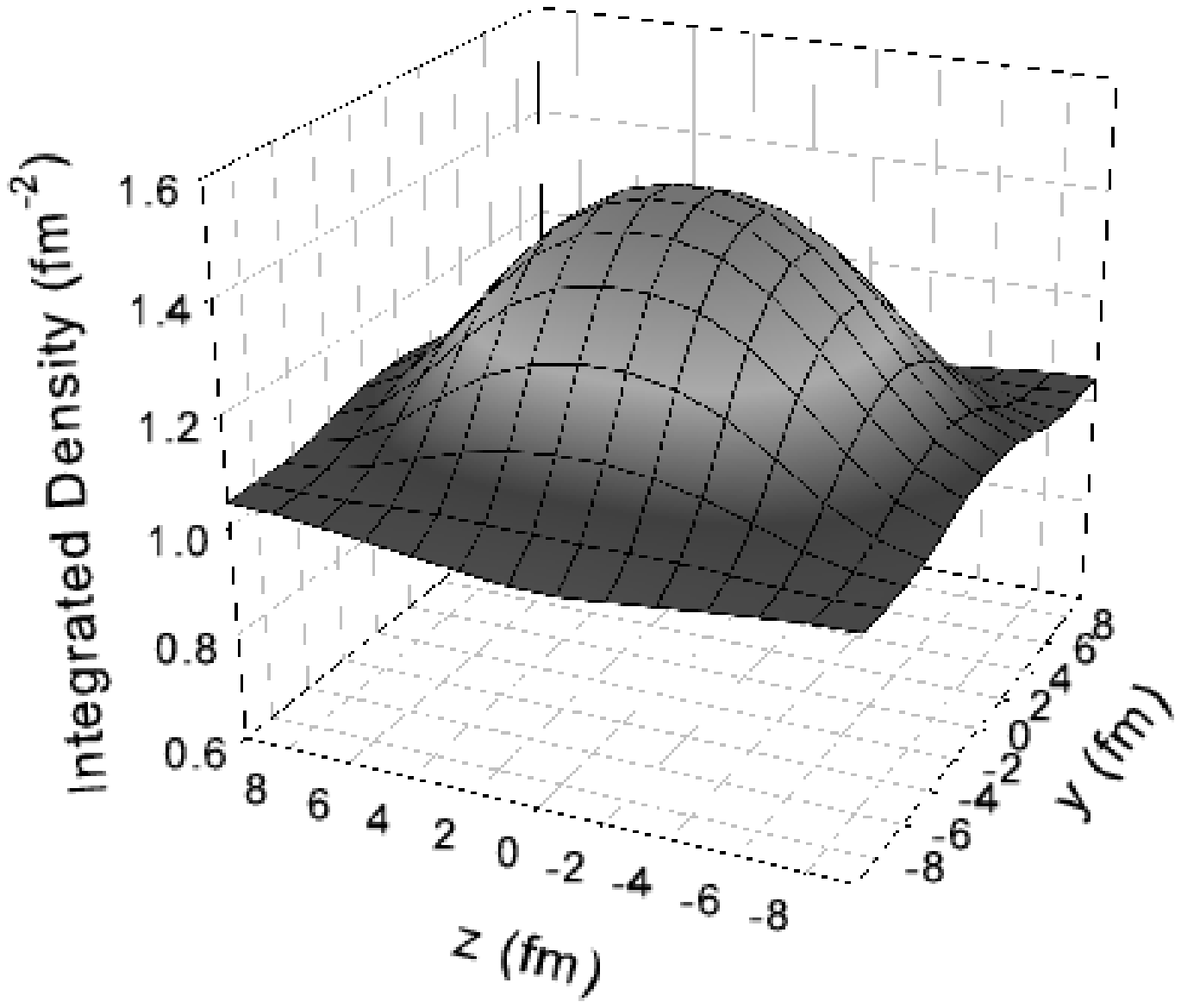,width=5.3cm}\psfig{file=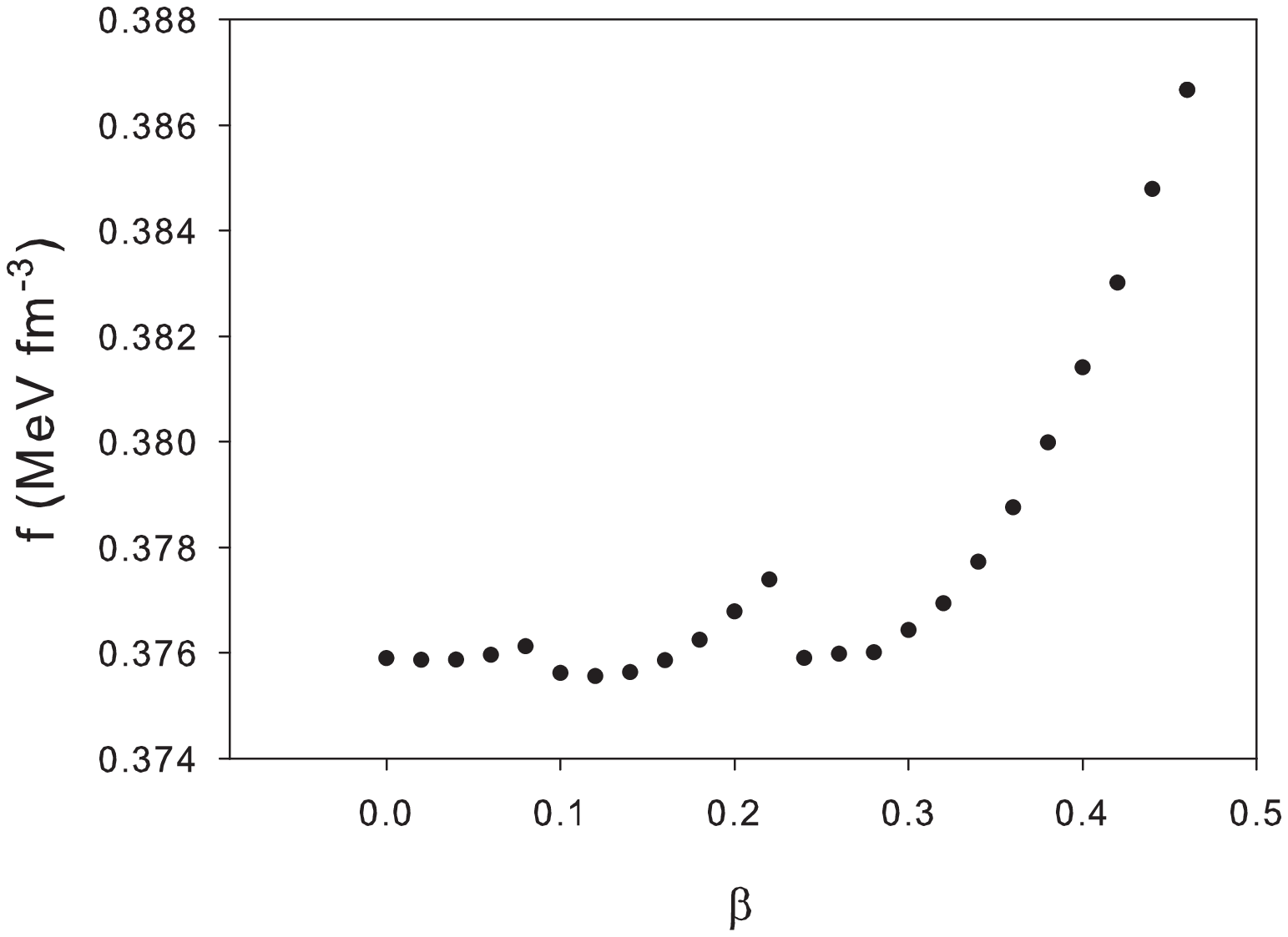,width=6.2cm}\psfig{file=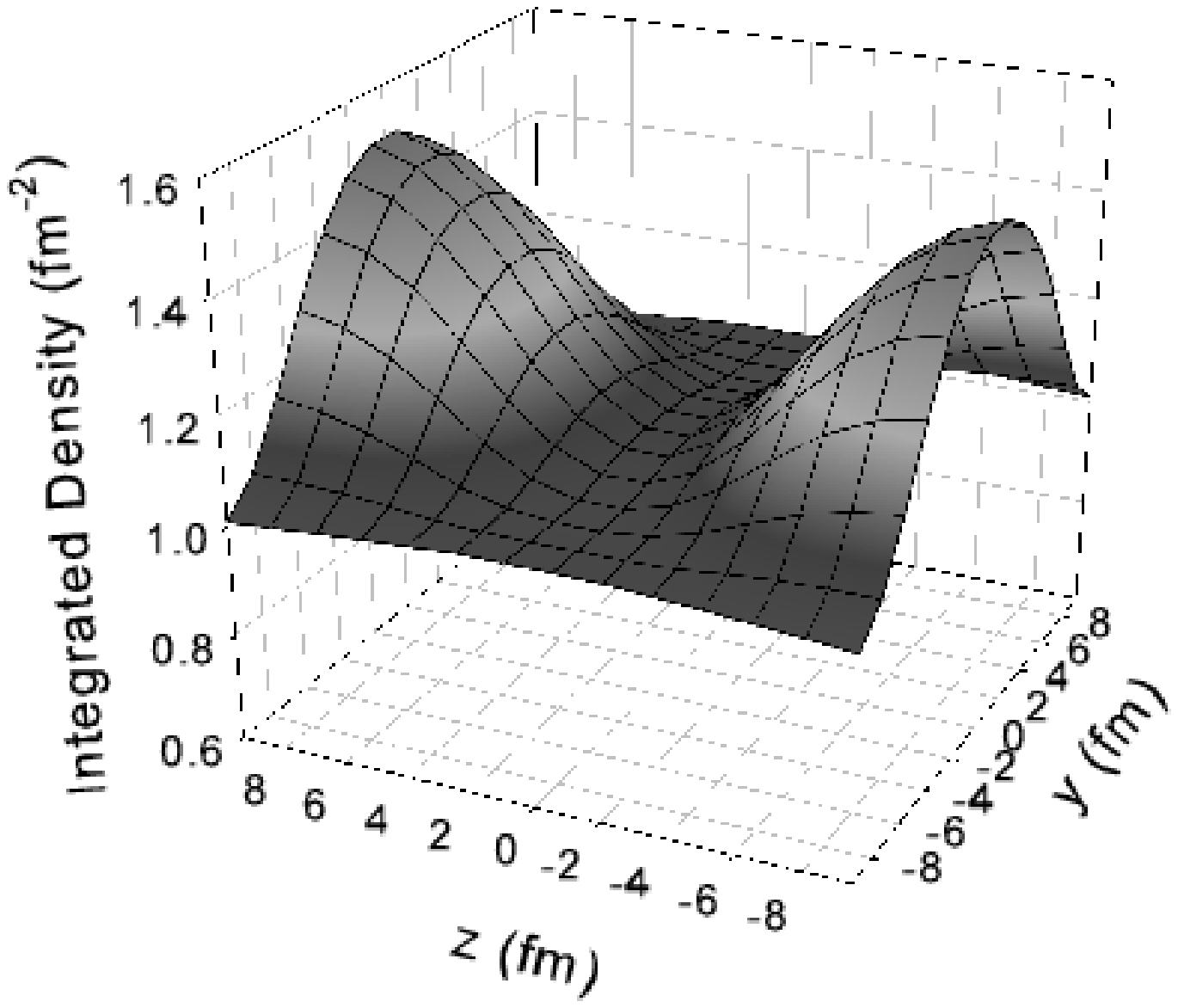,width=5.3cm}}
\caption{\label{Fig:2} Middle plot: Free energy density versus deformation parameter $\beta$, for $n_{\rm b}$ = 0.06 fm$^{\rm -3}$, $A$ = 500, $y_{\rm p}$ = 0.04, and $\gamma$ = 0$^{\rm o}$. Left and right: integrated density distributions integrated over the x-direction for $\beta$ = 0 and 0.24, $\gamma$ = 0$^{\rm o}$.} \hspace{1pc}
\end{figure}

The results of a second test are displayed in fig.~\ref{Fig:2}. The central panel shows the free energy density of a configuration as the deformation in the z-direction, $\beta$, is increased at $\gamma = 0^{\rm o}$. The two panels on either side display the density of the neutron distribution integrated over the x-direction for $\beta$ = 0.0 and 0.24. As the deformation $\beta$ increases, there are two sharp transitions. The nuclear configuration starts as a spherical nucleus immersed in a neutron gas, displayed on the left hand side. The first energy jump close to $\beta = 0.10$ corresponds to a transition from a deformed nuclear configuration to a cylindrical configuration - `spaghetti'. The second transition close to $\beta = 0.22$  is in fact a transition back to the original spherical configuration, except that the computational volume is now centered halfway between adjacent nuclear configurations, rather than on the configuration itself. The fact that the free energy density obtained for both configurations is identical indicates that the finite cell size does not introduce a spurious energy.

\paragraph{Spurious Shell Effects}

\begin{figure}[!t]
\hspace{1pc}
\centerline{\psfig{file=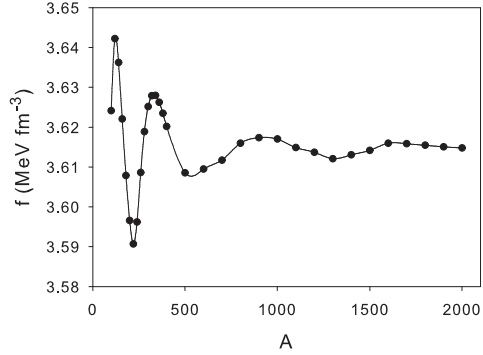,width=7cm}}
\caption{Free energy-density $f$ versus nucleon number $A$ at $n_{\rm b}$ = 0.11 fm$^{\rm -3}$, ($\beta,\gamma$)=(0,0$^{\rm o}$) and T = 2.5 MeV. The form of the curve is dominated by a spurious shell energy caused by the finite box discretization of the continuum neutron energy spectrum.} \hspace{1pc} \label{Fig:3}
\end{figure}

Although the nuclear structures obtained are not artifacts of the finite cell size, the finite cell size can be a source of spurious shell effects. In analogy with a Fermi gas in a box, shell effects caused by the discretization of the physical space due to the finite computational volume may occur. These effects will manifest themselves especially at high densities and temperatures when a large number the nucleons are unbound. To illustrate the form of these numerical shell effects, the free energy-density is plotted in Fig.~\ref{Fig:3} as a function of $A$ in a wide region up to $A$=2000 at a density of $n_{\rm b}$ = 0.11 fm$^{\rm -3}$, $(\beta, \gamma)$ = (0.0,0$^{\rm o}$) and $T$ = 2.5 MeV. The pronounced oscillations in the free energy-density curve are entirely numerical in origin. With increasing box size and temperature the bulk energy-density of the matter in the box approaches that derived from analytic formulae for the free energy-density of a free nucleon gas. These numerically induced shell effects manifest themselves in a form distinct from the physical shell effects, arising due to a combination of the shell energies of bound nucleons and unbound neutrons scattered by the bound nucleons, which are characterized by more rapid fluctuations in nucleon number $A$ typically at lower densities and temperatures and lower values of $A$. The distinction between the form and occurrence of the two types of shell effects is encouraging as it allows their easy identification. In such a situation where the shell effects are purely spurious, the physical value of the free energy-density is not the minimum, but that value to which the free energy tends at high $A$.

In order to accurately locate free energy-density minima as a function of $A$ and in energy-density surfaces, the numerical shell effects, which introduce spurious minima, should be identified and corrected for, a work currently in progress.

\subsection{Coverage of Parameter Space}

For each run of the code at a given temperature $T$ and baryon number density $n_{\rm b}$ = $A/V$, where $A = N + Z$ is the number of nucleons ($N$ neutrons and $Z$ protons), and $V$ the volume of the unit cell, up to four free parameters need to be specified: $A$, the proton fraction $y_{\rm p}$ = $Z/A$, and the quadrupole moment parameters $\beta$ and $\gamma$. Clearly at a given value of $n_{\rm b}$, $A$ and $V$ are not uniquely determined and calculations have to be performed over a range of values of $A$, or equivalently $V$, at that given density to find the value which gives the minimum free energy-density. The proton fraction is kept constant at $y_{\rm p}=0.3$ for SN matter calculations \cite{lattimer91}, whereas for NS matter the proton number must be varied to determine the energy minimum with respect to $y_{\rm p}$ corresponding to beta-equilibrium. Finally, when the quadrupole constraint is applied, the deformation parameters $\beta$ and $\gamma$ are also free parameters and must be varied over the deformation space, searching for shapes corresponding to minima of the total energy.

To build an EoS of the pasta phases of inhomogeneous matter, such a series of calculations has to be performed at each point of the parameter space ($n_{\rm b}, T, y_p$) with adequate spacing, which is extremely computationally intensive. Part of the task is to find the optimum strategy to obtain adequate coverage of parameter space; for example, some areas of parameter space are physically less important than others; this will lead to selection of those values of the parameters that are most likely to be physically manifest.

\begin{figure}[!t]
\hspace{1pc}
\centerline{\psfig{file=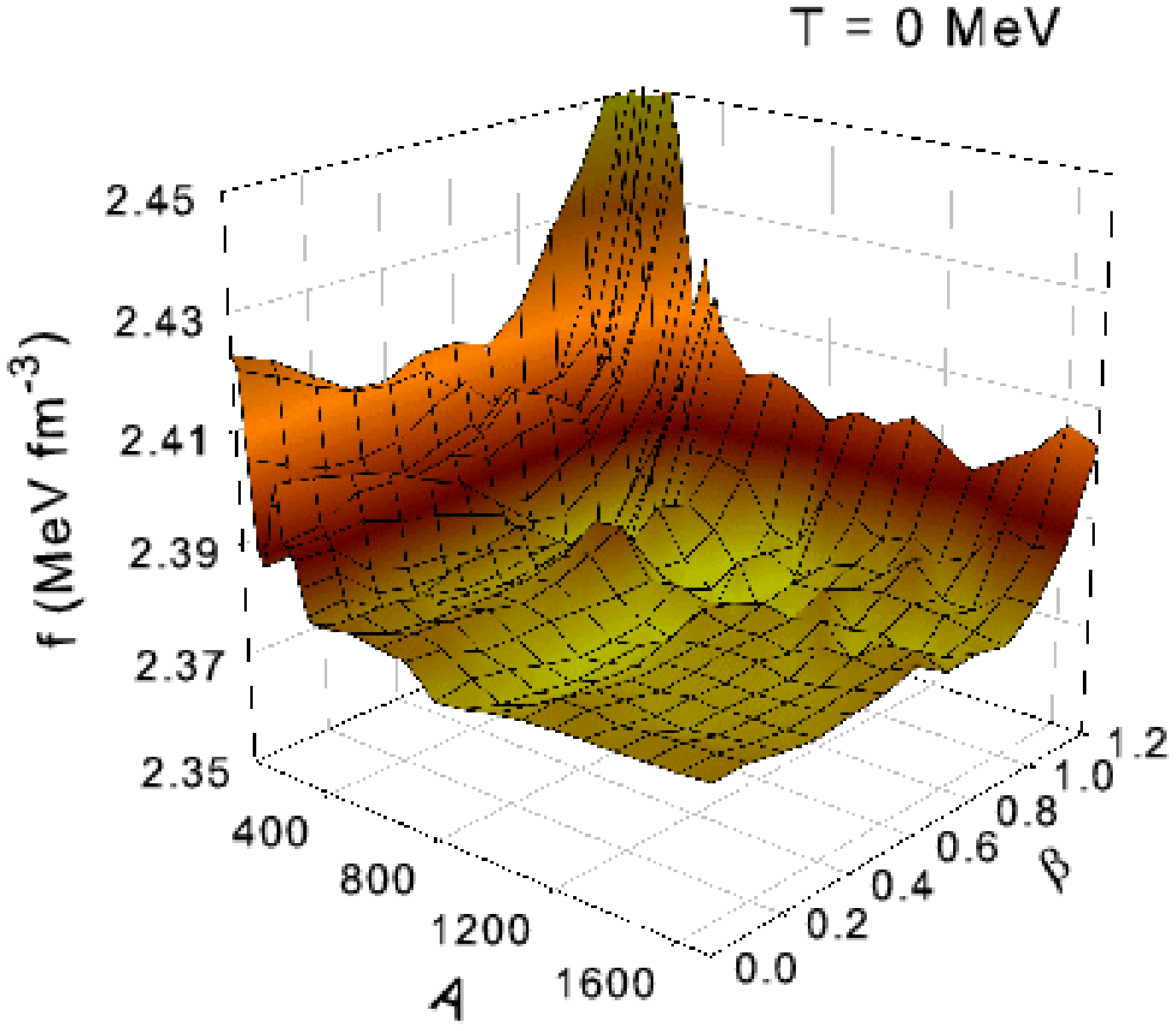,width=5.8cm}\psfig{file=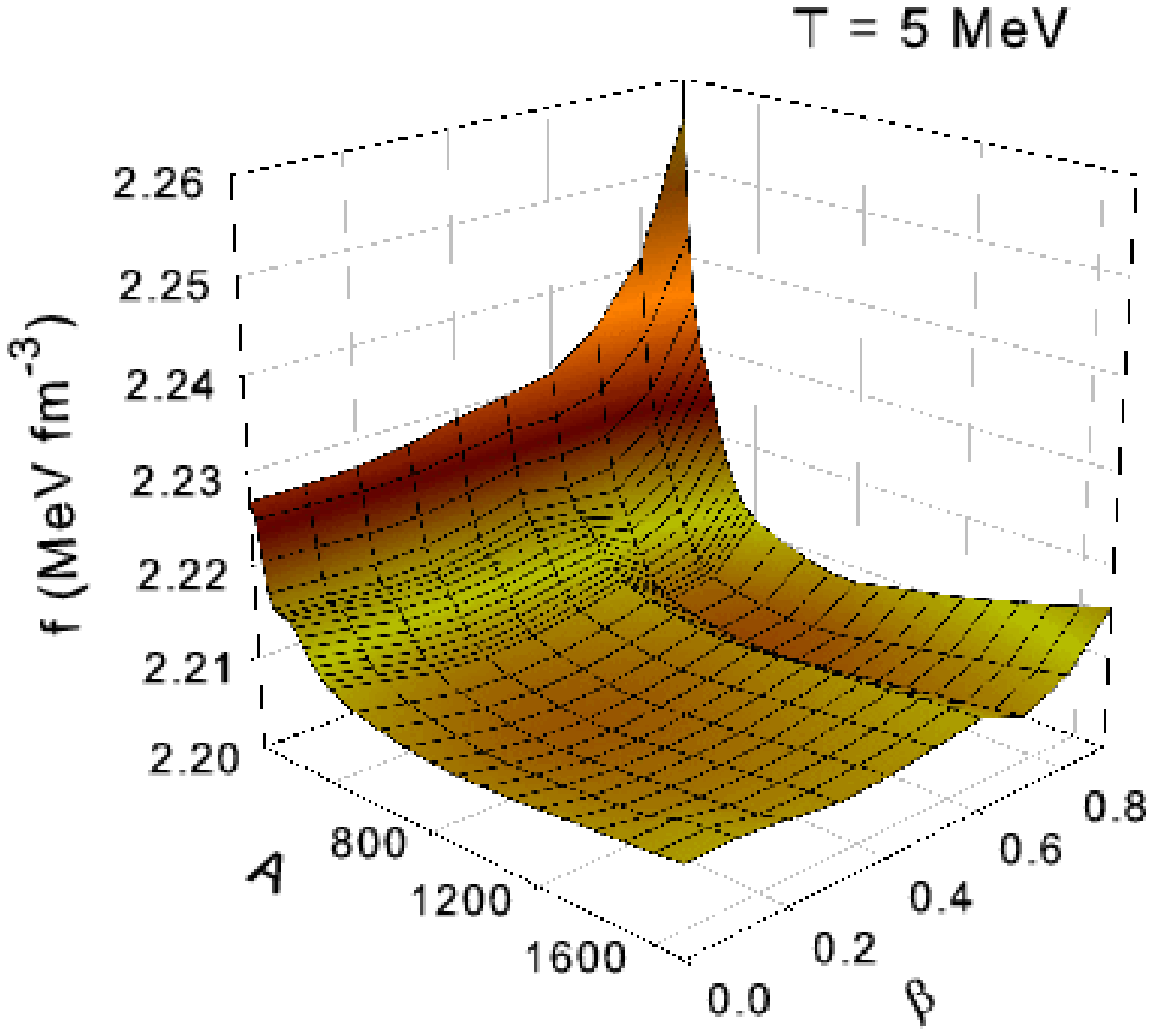,width=5.8cm}}
\caption{Free energy-deformation surfaces at n$_{\rm b}$ = 0.08 fm$^{\rm -3}$, $\gamma$ = $60^{\rm o}$ and $T = 0$ MeV (left) and $T = 5.0$ MeV (right).} \hspace{1pc} \label{Fig:4}
\end{figure}

In the case of SN matter, for each density calculations were performed at $y_{\rm p}$ = 0.3, $T$ = 0, 2.5, 5, 7.5 MeV, values of nucleon number $A$ = 100-3000 with a step of 100, $\beta$ = 0.0  - 2.0 and $\gamma$ = 0$^{\rm o}$, 30$^{\rm o}$ and 60$^{\rm o}$. Fig.~\ref{Fig:4} displays the energy-deformation surfaces obtained at the density values n$_{\rm b}$ = 0.08 fm$^{\rm -3}$ and $\gamma$ = 60$^{\rm o}$ for temperatures of $T$ = 0 MeV and 2.5 MeV. Physical shell effects associated with bound nucleons are visible at zero temperatures. The result of the shell effects is many local minima in the surfaces at zero or low temperature, of order of 5 keV fm$^{\rm -3}$ deep. The variation in shell energy with respect to $\beta$ at constant $A$ is expected to be caused by the varying shell structure of the bound nucleons. Magierski and Heenen~\cite{magie02} suggest that in \emph{neutron star} matter, the shell effects of the unbound neutrons (caused by their scattering off the nuclear structure) contribute significantly to the variation of energy with deformation; however, this is not expected to be the case in SN matter because the density of unbound neutrons is an order of magnitude smaller than in NS matter at a corresponding density.

The shell effects become washed out at higher temperature leaving behind the minima corresponding to different nuclear geometries - minima that result from the competition of nuclear surface energy and Coulomb energy. These occur not only with respect to $A$, but also with respect to $\beta$, and result in several local geometrical minima appearing on each energy-deformation surface. For example, at $T$ = 5.0 MeV there are two broad minima that appear, centered at $\beta \approx 0.3, A \approx 800$ and $\beta \approx 0.7, A \approx 800$. The geometrical configurations they correspond to are slab like and cylindrical hole like respectively. They are separated by a `ridge' in the surface, running along $\beta = 0.6$ which delineates the regions on the energy surface where the two geometries occur. The free energy-density changes smoothly in the vicinity of the minima, and the nuclear geometry locally remains the same.

\begin{figure}
\centerline{\psfig{file=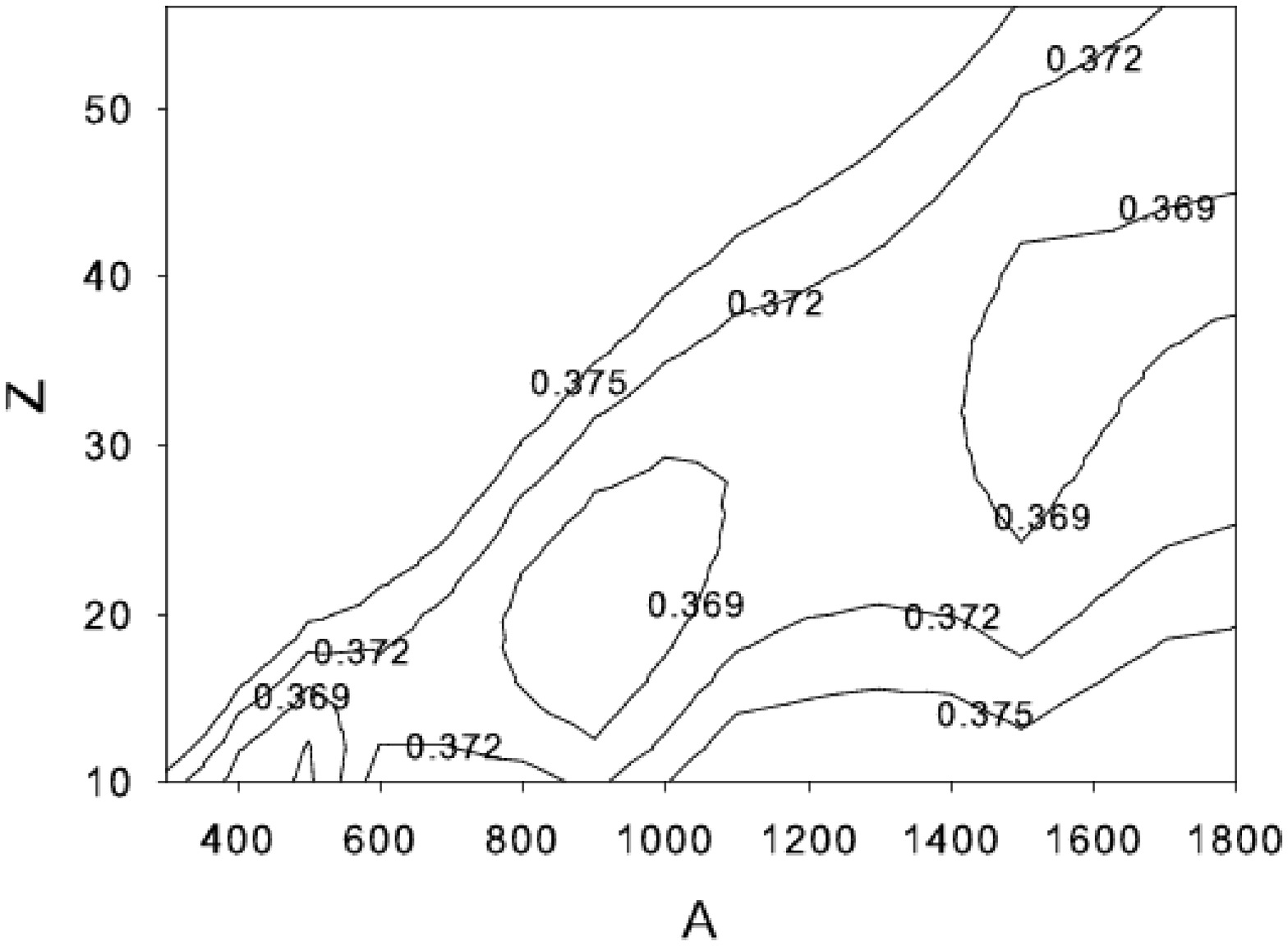,width = 7.5cm}\psfig{file=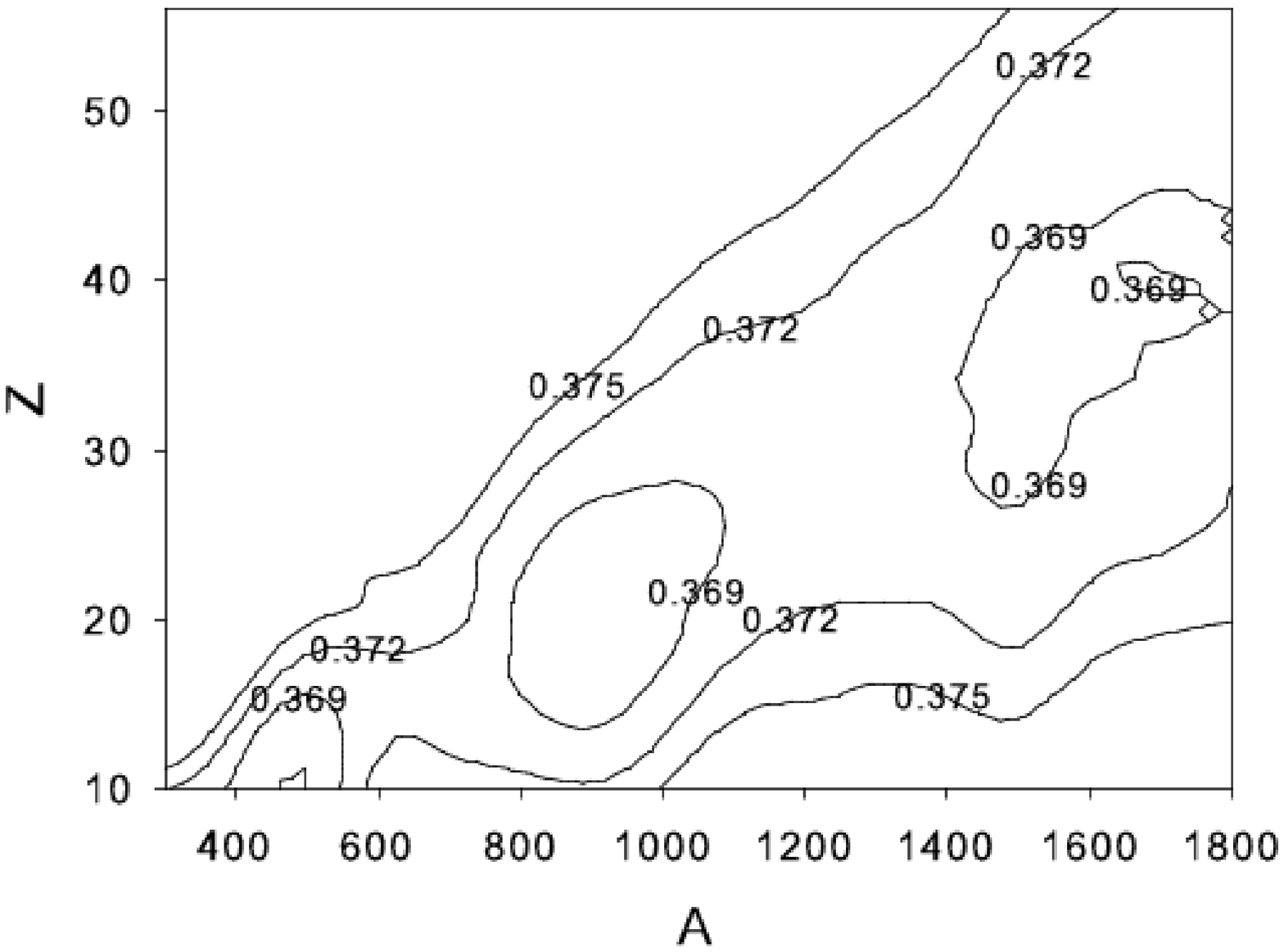,width = 7.5cm}}
\caption{Free energy density contour plots as a function of $Z$ and $A$ for $n_{\rm b}$ = 0.06 fm$^{\rm -3}$, $\gamma = 0^{\rm o}$ and $\beta = 0.0$ (left) and $\beta = 0.12$ (right).}
\label{Fig:5}
\end{figure}

\begin{figure}[!t]
\centerline{\psfig{file=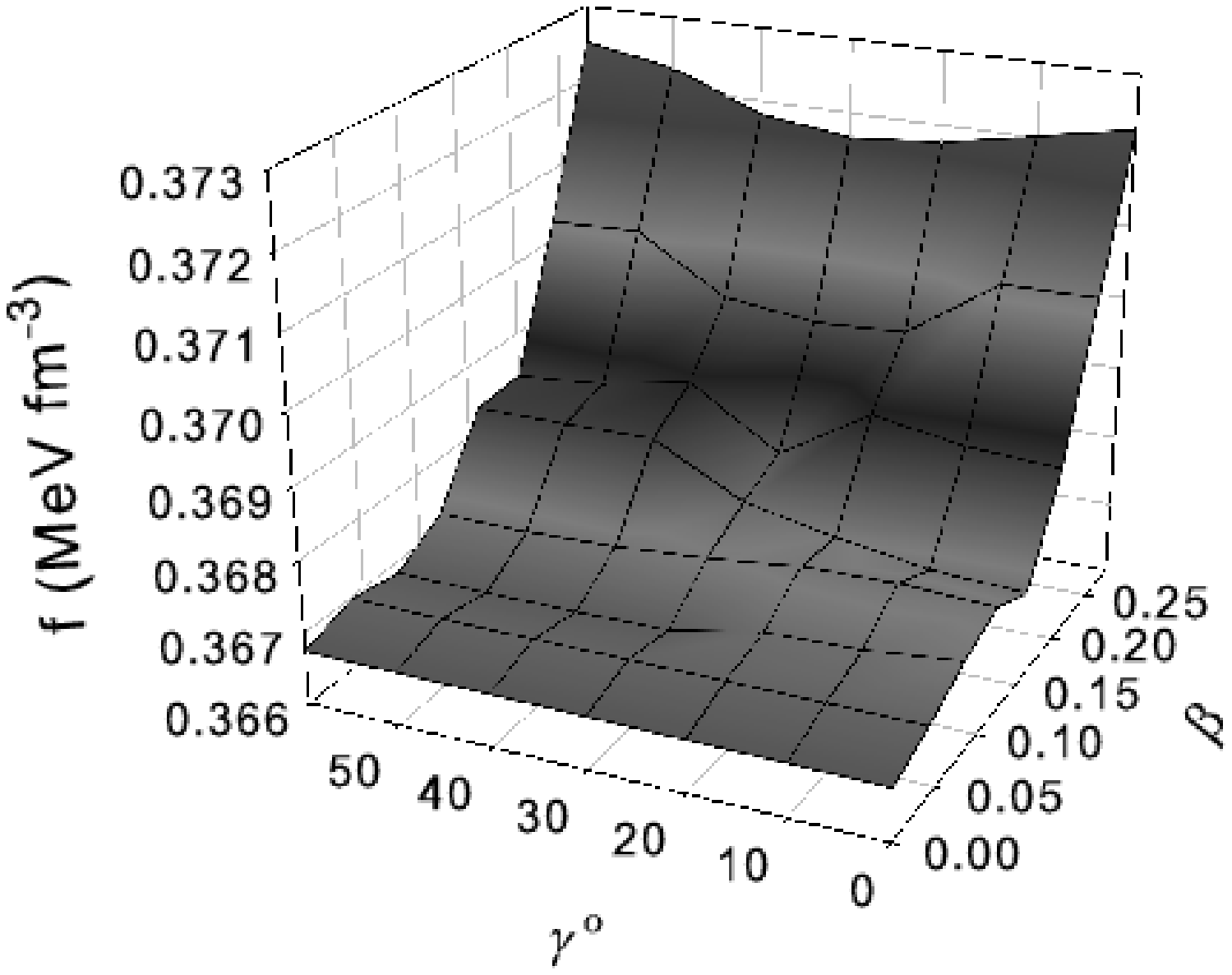,width = 6.5cm}\psfig{file=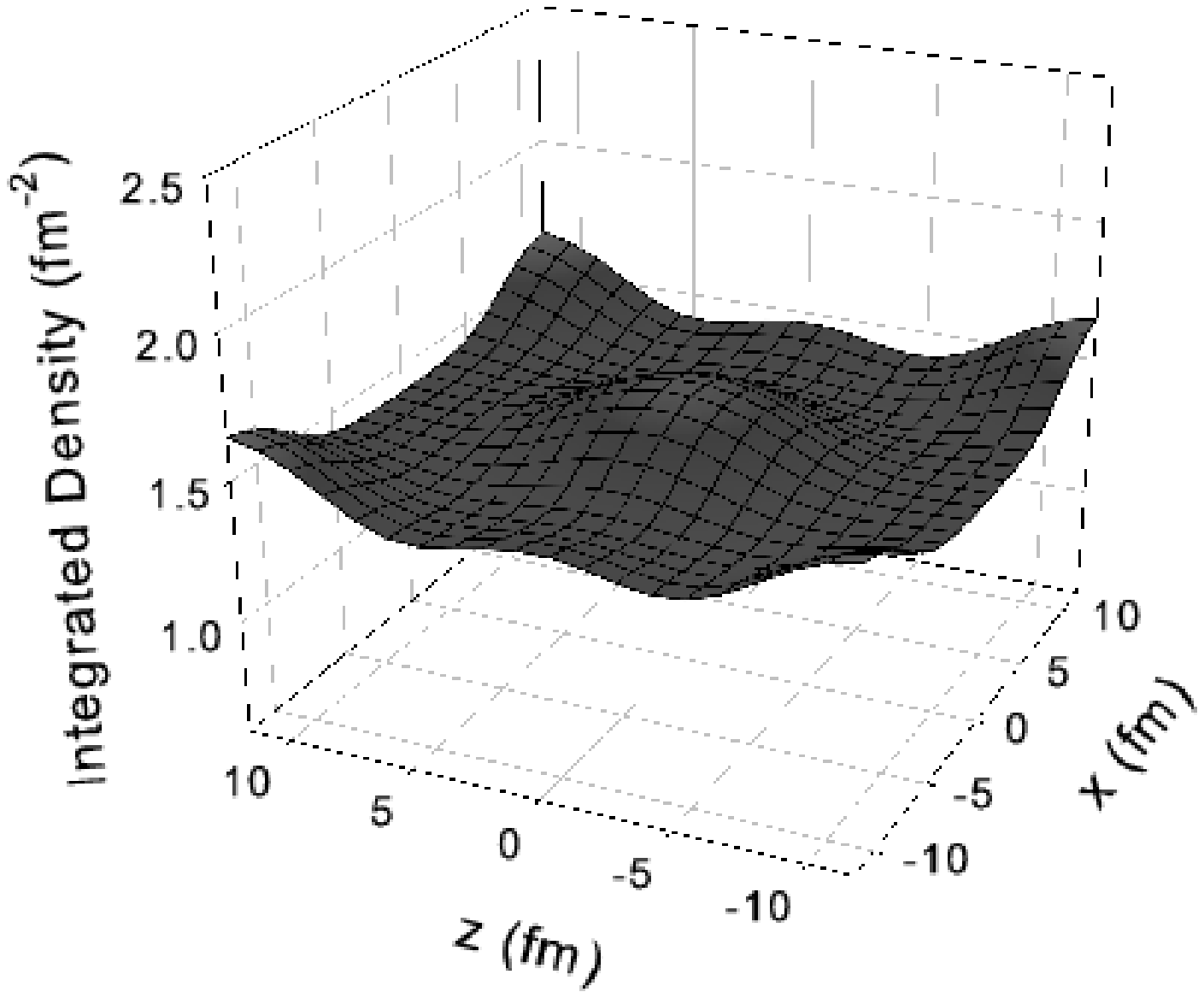,width = 7cm}}
\caption{Left: Free energy density versus deformation parameters $(\beta, \gamma)$ for minima at, $(A,Z)$ = (900,20). Right: z-integrated neutron density distributions for the absolute minimum energy configuration.}
\label{Fig:6}
\end{figure}

An example of a calculation of the preferred state of NS matter is given in Fig.~\ref{Fig:5}-~\ref{Fig:6}. The density is $n_{\rm b}$ = 0.06fm$^{-3}$. To obtain configurations of beta-equilibrium matter in the inner crust of NS, the free energy density at a given deformation $(\beta, \gamma)$, must be not only minimised with respect to the total nucleon number $A$ but also with respect to the proton number $Z$. This procedure is illustrated in Fig.~\ref{Fig:5} at a density of $n_{\rm b}$ = 0.06 fm$^{\rm -3}$, $\gamma = 0^{\rm o}$ and for two different values of $\beta$: 0.0 and 0.12. It is generally found that the position of the minimum with respect to $A$ and $Z$ is independent of $(\beta, \gamma)$. Thus the $(A,Z)$ position of minima at a zero deformation can be obtained, and then a search over deformation space at those specific $(A,Z)$ values gives the minimum energy nuclear shape. The left hand panel of Fig.~\ref{Fig:6} shows the energy as a function of deformation at $(A,Z) = (900,20)$. The right hand panel shows the absolute minimum energy configuration in the form of the neutron density integrated over the y-direction; in this case, it is a bcc lattice of roughly spherical nuclear shapes

\subsection{The Pasta Shapes}

A wide range of nuclear configurations emerge from out calculations. At $y_{\rm p} = 0.3$ and at temperatures up to $T$= 5 MeV, the canonical `pasta' shapes appear as absolute minima in the total energy for both Skyrme parameterizations used at densities of $\sim$0.04 fm$^{-3}$ (spherical nuclei), $\sim$0.06 fm$^{-3}$ (cylindrical nuclei), $\sim$0.08 fm$^{-3}$ (slab nuclei), $\sim$0.09 fm$^{-3}$ (cylindrical holes in nuclear matter), $\sim$0.10 fm$^{-3}$ (spherical holes in nuclear matter), with uniform matter appearing at higher densities.

\begin{figure} [!t]
\centerline{\psfig{file=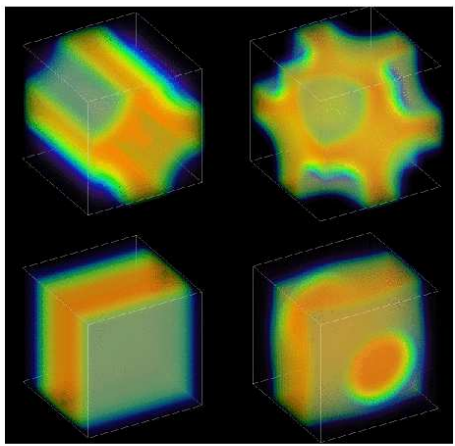, width=7.5cm}} \caption{Nuclear configurations obtained at local minima at $n_{\rm b}$ = 0.083 fm$^{-3}$, $y_{\rm p}$ = 0.3 and $T$ = 2.5MeV: top left - A=460, top right - A=2200, bottom left - A=280, bottom right - A=1300.} \label{Fig:7}
\end{figure}

However, when one looks at the energy surface at a given density, many of the pasta shapes appear as local minima separated from the absolute minima by up to a few keV fm$^{-3}$, suggesting that matter is somewhat disordered with no one phase of nuclear shape existing over a macroscopic extent. Fig.~\ref{Fig:7} shows renderings of the neutron density distribution at $n_{\rm b}$ = 0.083 fm$^{-3}$, $y_{\rm p}$ = 0.3 and $T$ = 2.5MeV obtained in various local minima in the energy-deformation surface. Slab nuclei, cylindrical holes and spherical holes and intermediate pasta shapes all appear in local minima separated from the absolute minima by less than 5 keV fm$^{-3}$.

Finally, many shapes different to the canonical pasta shapes are observed, often as local minima themselves. It is likely that the transition between the canonical pasta shapes is mediated by a series of transitions through intermediate shapes. All this supports the argument that the the thermodynamic quantities of bulk inhomogeneous matter change smoothly through the pasta region up until the final phase transition to uniform matter, which is weakly first order.

\subsection{The Nuclear Force Dependence on the Appearance of Pasta}

\begin{figure}[!t]
\centerline{\psfig{file=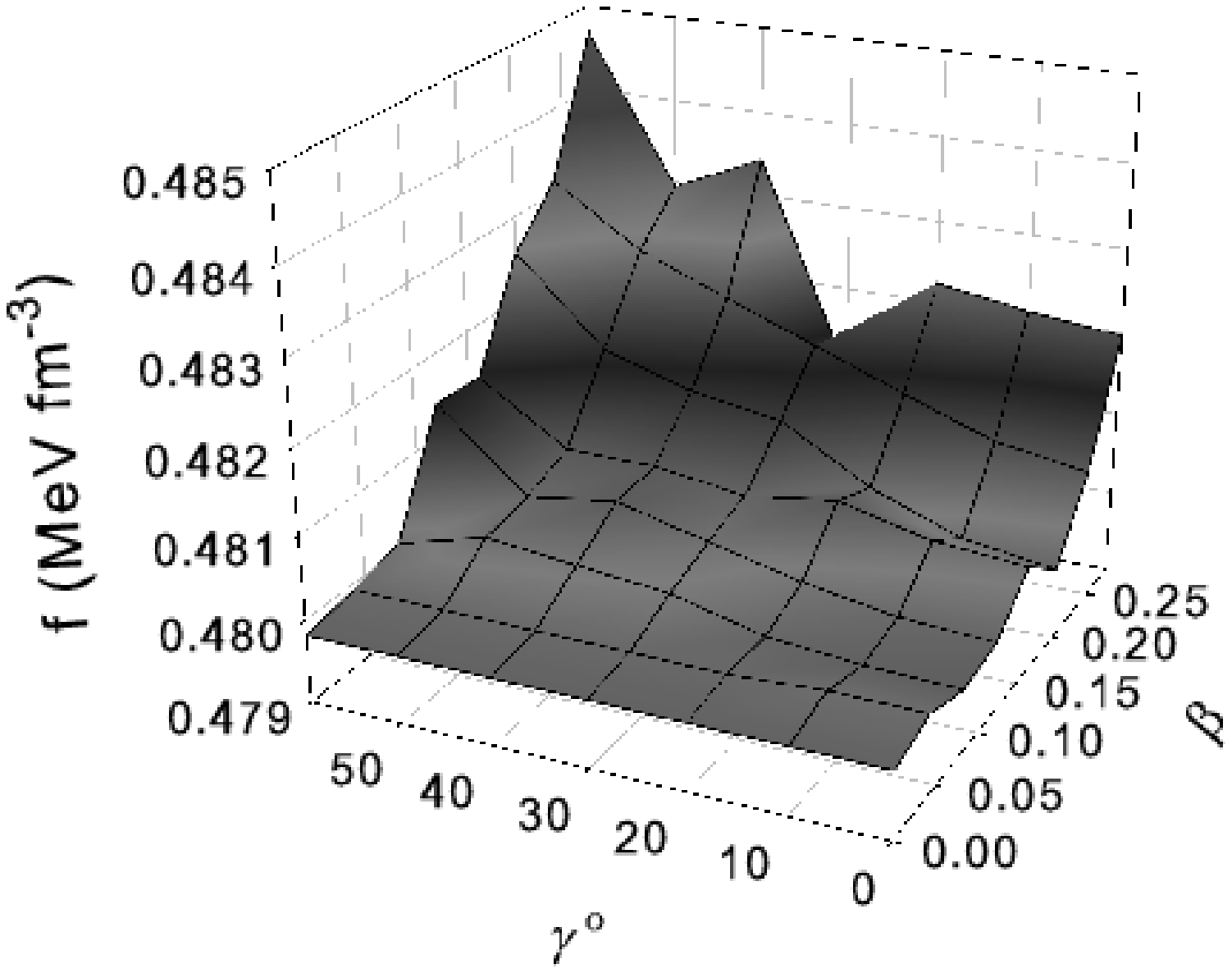,width = 6.5cm}\psfig{file=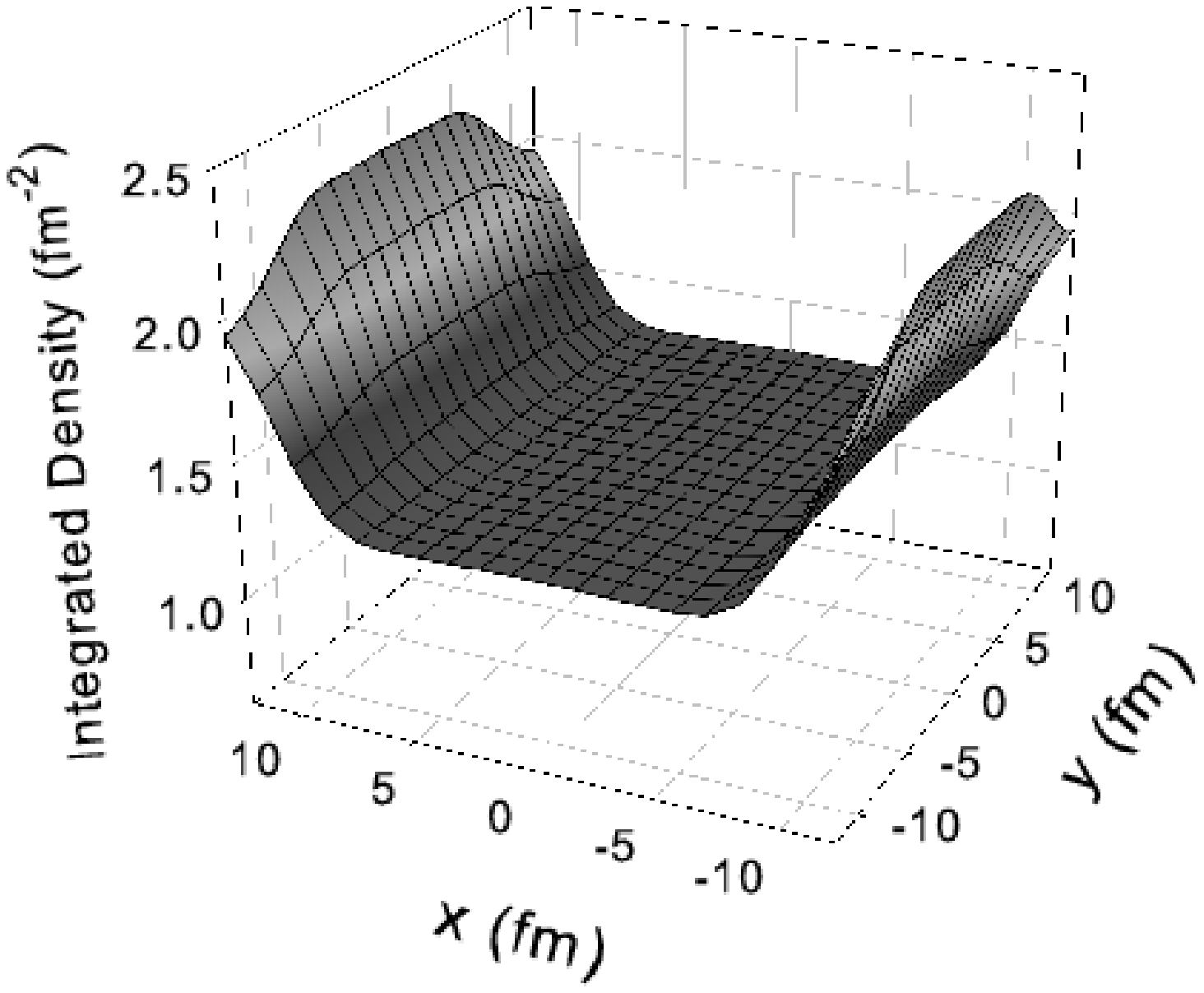,width = 7cm}}
\caption{Left: Free energy density versus deformation parameters $(\beta, \gamma)$ for the minimum at $(A,Z)$ = (900,30). Right: z-integrated neutron density distribution for the absolute energy minimum.}
\label{Fig:8}
\end{figure}

The dependence of the density regions in which various pasta shapes appear on the Skyrme parameterization is currently being tested; a preliminary result is given here. The NS matter calculation displayed in Fig.~\ref{Fig:5}-~\ref{Fig:6} is repeated for the Sly4 parameterization \cite{Chaba1998}
Similarly to Fig.~\ref{Fig:6}, the left hand panel of Fig.~\ref{Fig:8} shows the energy as a function of deformation for Sly4 at $(A,Z) = (900,20)$ (one of the beta-equilibrium minima for this parameterization). The right hand panel shows the absolute minimum energy configuration in the form of the neutron density integrated over the z-direction. Here, a slab-like nuclear shape is obtained as opposed to the absolute minimum nuclear shape for SkM$^*$, which was spherical. Indeed, all minima found so far using the SkM$^*$ \emph{at beta-equilibirum} have corresponded to roughly spherical nuclear distributions - that is, no pasta shapes have so far been obtained as absolute minima at a given density. The Sly4 parameterization results in a rich variety of pasta shapes at energy minima at beta-equilibrium, in a sharp contrast to the SkM$^*$ parameterization. The crucial difference between the two parameterizations is the higher symmetry energy of Sly4 over the pasta density range, which gives a larger proton fraction at beta-equilibrium. A higher proton fraction promotes larger nucleon clusters, and pasta shapes become more energetically favorable. Indeed, at higher proton fractions, pasta shapes appear for SkM$^*$ as energy minima with respect to $\beta$, but of course these configurations are no longer in beta-equilibrium. This echoes the results of studies which use semi-classical models; see, for example, the dependence of the crust-core transition on asymmetry energy \cite{Steiner2008}. Also, this result is at odds with some semi-classical models which use the Sly family forces and yet find that pasta phases are not energetically preferred \cite{Douchin2001}. No pasta in the crust would lead to a sharper phase transition to uniform matter and a clear delineation of the solid crust and liquid core. Pasta in crust would imply a smoother, more gradual phase transition to uniform matter as well as a more gradual transition of the mechanical (and other) properties of the crust. The difference between these two scenarios may translate into a prediction of an observable difference in the dynamics of neutron stars and a potential probe of the asymmetry energy of sub-nuclear matter. Investigation of this feature over a wider range of Skyrme forces is in progress.

One final note: a preliminary investigation of SN matter, where the proton fraction is held fixed at $y_{\rm p}$ = 0.3 independent of Skyrme used, indicates a much weaker dependence of the pasta phase diagram on choice of parameterization.

\subsection{The Transition to Uniform Matter}

By taking the minimum free energy-density configurations of the densities and temperatures for which  SN matter was calculated ($n_{\rm b}$ = 0.04, 0.06, 0.08, 0.09, 0.10, 0.11, 0.12 fm$^{-3}$, $T$ = 0, 2.5 and 5 MeV) the EoS can begin to be constructed.

\begin{figure}[!t]
\centerline{\psfig{file=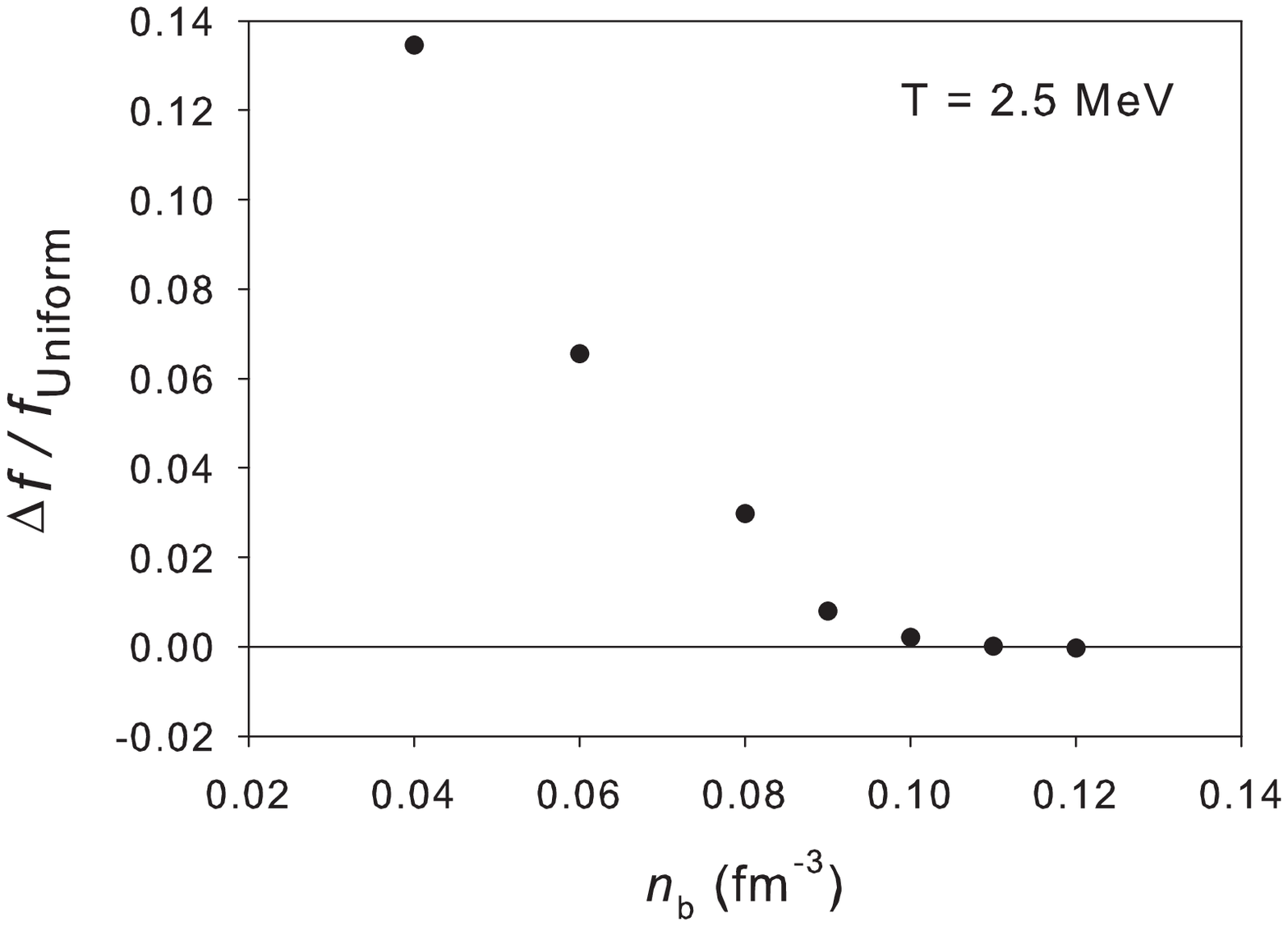,width = 9.0cm}\psfig{file=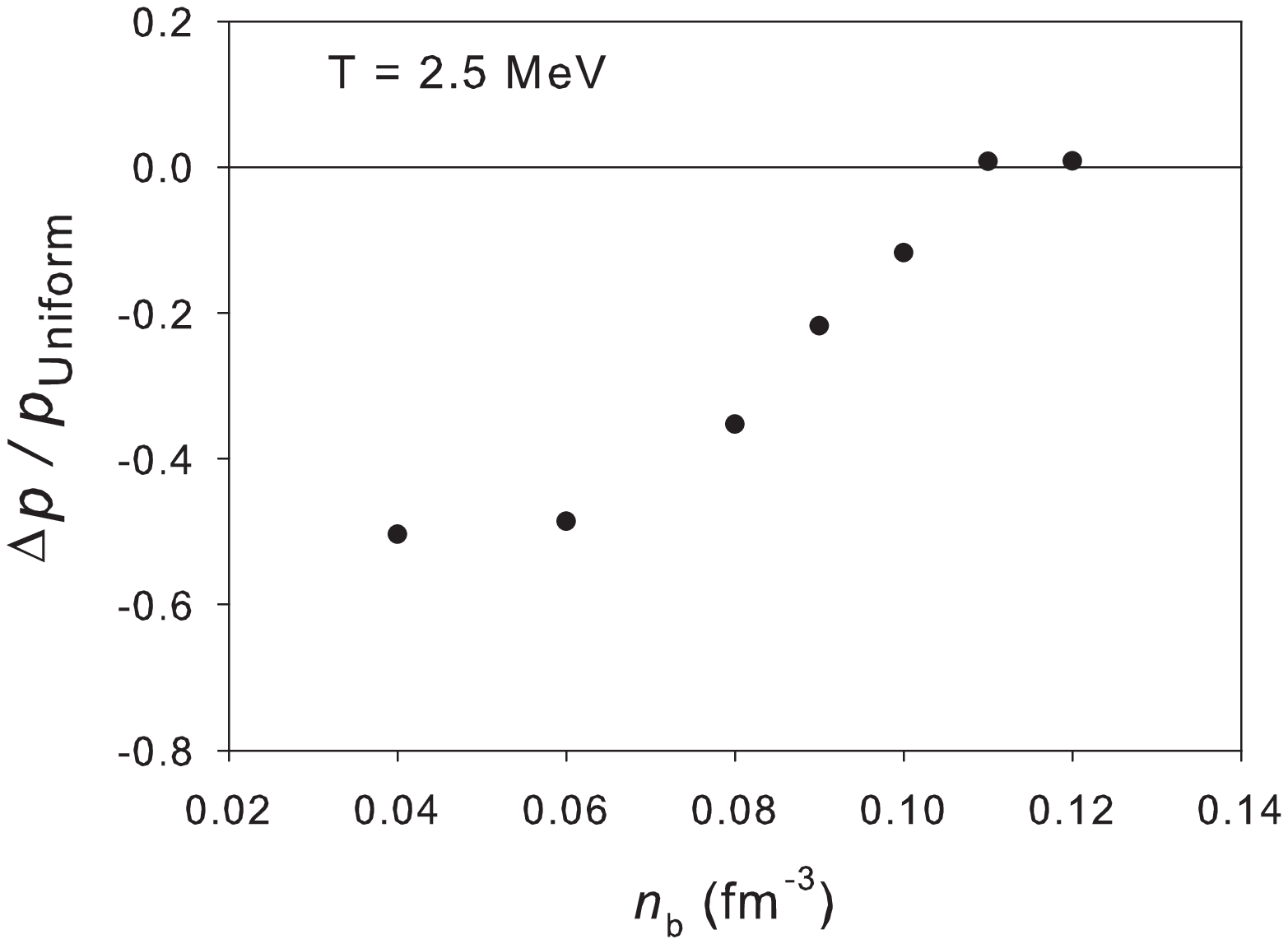,width = 9.0cm}}
\caption{Left: Difference between the free energy-density of uniform matter and non-uniform matter $\Delta f$ divided by the free energy-density of non-uniform matter $f_{\rm Uniform}$ as a function of density at $T$ = 2.5 MeV. Right: Difference between the pressure of uniform matter and non-uniform matter $\Delta p$ divided by the pressure of non-uniform matter $p_{\rm Uniform}$ at the same temperature.}
\label{Fig:9}
\end{figure}

For the present model to produce the transition to uniform matter consistently, the various thermodynamic quantities calculated should tend in some way towards those calculate in the uniform matter where the Skyrme energy-density is given analytically. The difference in free energy-density and pressure between the inhomogeneous matter and the uniform matter Skyrme-HF model is plotted in Fig.~\ref{Fig:9} for the densities listed above and a temperature of 2.5 MeV. The difference in free energy-density $\Delta f / f_{\rm Uniform} = (f_{\rm Uniform} - f_{\rm Non-Uniform})/f_{\rm Uniform}$ is displayed in the left plot and is seen to vanish relatively smoothly as the transition density $n_{\rm b} \approx 0.10 - 0.11$ fm$^{-3}$ is approached. The equivalent pressure difference $\Delta p / p_{\rm Uniform} = (p_{\rm Uniform} - p_{\rm Non-Uniform})/p_{\rm Uniform}$ is shown on the right hand plot and exhibits a discontinuity at the transition density in a way that suggests the occurrence of a phase transition, as expected. A similar discontinuity shows up when one looks at the variation of entropy with density at finite temperature. More detailed calculations around the transition density are required to determine the exact nature of the transition. Physically, the transition is expected to be first order, as in a liquid-gas transition.

One caveat; one should bear in mind that spurious shell effects, yet to be fully taken into account, might alter the position of the minima in parameter space at densities and temperatures close the transition to uniform matter.

\subsection{Ordering a Disordered Phase}

Our calculations, like recent 3D Hartree-Fock calculations, have indicated that the pasta phase is somewhat disordered: the large number of closely spaced local minima make it likely that a variety of phases will coexist at a certain density. In that case, one can expect an averaging out of their mechanical and thermal properties on a meso- and macroscopic scale. Is there any way of ordering the phases? One possibility is the magnetic field: here the energy density of the magnetic field is estimated and compared to the typical difference between minima. For a typical \emph{surface} magnetic field of an ordinary radio pulsar, $B \approx 10^{12}$ G, the energy density of the field is $E_B = {1 \over 2 \mu_0} B^2 \approx 10^{-5}$ keV fm$^{-3}$. This is much less than the typical difference between minima $\approx$~1-10 keV fm$^{-3}$. However, if one considers magnetic fields typical to magnetars $\approx 10^{15}$ G, and energy density of $E_B \approx 10$ keV fm$^{-3}$ is obtained. Magnetic fields of this order of magnitude could thus have an effect on the microscopic structure of matter. (As a final note, these are magnetic fields at the surface of the star; the magnitudes of fields deep in the crust are still a matter of speculation; however, they could reach magnitudes of $10^{16} - 10^{17}$ G).

\section{Conclusions and Future Development}

The self-consistent 3D Skyrme-Hartree-Fock+BCS method has been applied to the study of the inhomogeneous `pasta' phase of bulk nuclear matter. Constraints on both independent quadrupole moments of the neutron density $(\beta, \gamma)$ allow a self-consistent determination of the energy-density of different nuclear geometries at a given density and temperature. Nucleon number $A$ and, in neutron star matter proton number $Z$, are also free parameters. The variation of all free parameters at a given density and temperature maps out an energy-deformation surface $f = f(A, Z, \beta, \gamma)$. It has been demonstrated that the present model smoothly spans the transition from isolated, roughly spherical nuclei surrounded by a low density gas of nucleons to uniform nuclear matter with increasing temperature and baryon number density. The free energy-density of non-uniform matter smoothly tends toward that of uniform matter as the transition density is reached. Discontinuities in the pressure and entropy are suggestive of a phase transition to uniform matter; further calculations are required to determine whether that phase transition is first or second order. The five canonical types of pasta are seen to emerge naturally from the model as well as a large variety of nuclear shapes intermediate between them. At a given density and temperature, many local minima occur corresponding to various nuclear geometries, indicating that a variety of pasta shapes coexist and the thermodynamical properties vary smoothly throughout the pasta regime rather than a series of phase transitions occurring from one bulk phase made up of one specific pasta shape to another: the properties of the different pasta configurations that make up the matter are likely to be averaged on a meso- and macroscopic scale.
Work is in progress on extending of the present calculation over a larger area of baryon number density, proton fraction and temperature using multiple Skymre parameterizations, on extending the current model to include the more general Bloch boundary conditions and so calculate the band structure of the pasta phases, and on calculating accurately the density of transition to uniform matter in NS and SN matter.


\begin{theacknowledgments}
The authors gratefully acknowledge Tony Mezzacappa for stimulating discussions, encouragement and support of access to the leadership computing facility in Oak Ridge, where majority of the calculations were done, and to NERSC computers at Lawrence Berkeley National Laboratory used during the initial phase. It is a pleasure to thank John Miller for continuous encouragment and comments during the course for this work, and to Chris Pethick and Philipp Podsiadlowski for interesting discussions and suggestions. Thanks also to P.-G. Reinhard, Raph Hix and Bronson Messer for helpful comments and to Constanca Providencia and Miguel Oliveira for supporting the final stage of the calculations at the super-computer Milipeia at Universidade de Coimbra, Portugal. An important part of the project was the visualization of the results which would not have been possible without substantial help of Chaoli Wang, Jonathan Edge, Amy Bonsor and Ross Toedte.

The major part of this work was conducted under the auspices of the TeraScale Supernova Initiative, funded by SciDAC grants from the DOE Office of Science High-Energy, Nuclear, and Advanced Scientific Computing Research Programs. Resources of the Center for Computational Sciences at Oak Ridge National Laboratory were used. Oak Ridge National Laboratory is managed by UT-Battelle, LLC, for the U.S. Department of Energy under contract DE-AC05-00OR22725. Partial support for this work by the UK EPSRC (WGN) and US DOE grant DE-FG02-94ER40834 (JRS) is acknowledged with thanks.
\end{theacknowledgments}

\end{document}